\documentstyle[aps,epsf,prl,twocolumn]{revtex}

\draft

\begin{document}

\title{Folding Langmuir Monolayers}

\author{Weixing Lu*, Charles M. Knobler*, and Robijn F. Bruinsma**}
\address{Departments of Chemistry and Biochemistry* and Physics**}
\address{University of California, Los Angeles, California 90095-1569}
\author{Michael Dennin and Michael Twardos}
\address{Department of Physics and Astronomy
and Institute for Interfacial and Surface Science}
\address{University of California, Irvine, CA 92697-4575.}

\date{\today}

\maketitle

\begin{abstract}

The maximum pressure a two-dimensional surfactant monolayer
is able to withstand is limited by the collapse instability
towards formation of three-dimensional material. We propose a
new description for reversible collapse based on a mathematical
analogy between the formation of folds in surfactant monolayers
and the formation of Griffith Cracks in solid plates under stress.
The description, which is tested in a combined microscopy and
rheology study of the collapse of a single-phase Langmuir monolayer
of 2-hydroxy-tetracosanoic acid (2-OH TCA), provides a connection
between the in-plane rheology of LM's and reversible folding.

\end{abstract}

\pacs{68.18.Jk,68.37.-d,83.60.Uv,87.68.+z}

The collapse of deformable structures under stress has fascinated
scientists and engineers since 1744 when Euler presented his
linear stability analysis of the large-scale structural failure of
plates, known as the {\it buckling instability} \cite{ref1}. In the
physics literature, recent research in this area has focused
on the local structure of crumpled surfaces \cite{ref2}
and on the geometry of the fold lines \cite{ref3}. Langmuir monolayers
(LM's), insoluble surfactant films resting on an air-
water interface, constitute a natural arena for the study of the
collapse of surfaces at the molecular level. LM's are stable at
low surface pressures $\pi$, but collapse into the third dimension
when sufficiently compressed. While
collapse can occur by the nucleation and growth of the bulk
phase, Euler-type
buckling phenomena are well documented \cite{ref4}.
Linear stability analysis \cite{ref5} shows
that when the surface tension $\gamma = \gamma_w - \pi$
of a LM vanishes (with $\gamma_w \approx 72\ {\rm mN/m}$
the surface tension of water), the air-water interface indeed should
become unstable against mechanical buckling. However,
the collapse of LM's exhibiting buckling-type phenomena actually
takes place for {\it positive} surface tensions $\gamma$
in the range of 1-10 mN/m, which casts doubt on Euler buckling as
the underlying mechanism. The introduction of spontaneous
curvature into the free energy of the film does lower the
instability threshold, but only by a small amount \cite{ref6}.

Recent experimental studies demonstrated that certain of these LM's
undergo {\it reversible} collapse, i.e., with little or no hysteresis in
the pressure-area isotherms. Although this was first seen under
conditions where there is coexistence in the LM between an isotropic
liquid phase and islands of a condensed phase \cite{ref7,ref8}
it has also been observed for a one-component monophasic
material \cite{ref9}. During reversible collapse, large isolated
folds into the subphase appear at the collapse point; they remain
fully connected to the monolayer and open reversibly when the
film is re-expanded. There is special interest in reversible
folding because of the recent proposal \cite{ref10} that it plays an
essential role in the functioning of lung surfactant.

The aim of this letter is to propose that reversible collapse in
LM's is unrelated to the (linear) Euler instability but
instead to a general, non-linear instability of {\it deformable
surfaces with self-attraction}. This instability is closely related
to a classical problem in structural failure of plates under
stress, namely crack formation. The ``mapping'' between the two
problems can be tested experimentally and provides a scenario
for reversible fold formation in terms of
the in-plane rheology of LM's.

The mechanism is shown in Fig.~1a. Assume a rectangular, self-adhering
elastic sheet of thickness $d$ with a two-dimensional (2D) shear
modulus $G$, an area modulus $K$, a bending modulus $\kappa$ and a
self-adhesion energy per unit area $W$. For LM's consisting of
long-tailed surfactant molecules, $W/2$ can be
identified as the free energy cost of creating a unit area of
hydrocarbon surface exposed to air, i.e., with the
{\it surface energy} $\gamma_H$ of a hydrocarbon fluid (for a
wide range of hydrocarbons $\gamma_H$ lies in the limited range of
$25 \pm 4\ {\rm mN/m}$ \cite{ref11}).

Let this sheet be stretched along the y direction by a traction
force per unit length $\gamma$ with the sheet width along the
x direction kept fixed. Now create a bilayer fold of length $L$
along the x direction, with a fold profile $U(x)$. This means
that the y-component of the in-plane elastic displacement field
undergoes a mathematical discontinuity $\pm U(x)$ along the fold
line. Let $\sigma_{ij}$ be the
contribution to the elastic stress tensor introduced by the
fold, so $\sigma_{ij} = 0$ far
from the fold. Under quasi-static conditions, the elastic
force per unit length $\gamma + \sigma_{yy}(x)$ pulling material
out of the fold should equal the force per unit length $W/2$
pulling material into the fold. Hence, $\sigma_{yy}(x)$
must be a constant along the fold line equal to $W/2 - \gamma$.

These unusual boundary conditions on the stress tensor have the same
mathematical form as those encountered in the materials-science
problem of the formation of cracks in plates under an external
traction force per unit area $P$, known as {\it Griffith Cracks}
(GC). Traction-induced crack formation of plates can
be considered as the analogue of adhesion-induced fold
formation of a LM, albeit with the sign of the displacement
field and the stresses reversed. For a
GC, the boundary condition along the crack surface states
that the total stress must vanish along the crack so
$\sigma_{yy} = -P$. By replacing $P$ with $\gamma - W/2$ in the
corresponding expressions for a Griffith Crack \cite{ref1}
we obtain the following prediction for the shape of a
fold in a solid LM:
\begin{equation}
U(x)/L = \frac{[W/2 - \gamma]}{Y}\sqrt{1 - (2x/L)^2}
\end{equation}
with $Y = 4KG/(K+G)$ the (two-dimensional) Young's Modulus.
Folds in solid LM's thus should have a {\it semi-elliptical} shape
with a ratio $\Gamma=U(0)/L$ between
short and long axes that should be a fold-independent
material parameter. According to Eq.~1, fold formation
indeed can start for positive surface tensions $\gamma$, i.e.
when the surface tension $\gamma$ drops below $W/2$, which
corresponds to surface pressures exceeding
$\gamma_w - \gamma_H = 53 \pm 4\ {\rm mN/m}$.

To test the proposed mapping, we examined folding in a relatively
simple system that involves only a single component  and a
single monolayer phase. The collapse of 2-hydroxytetracosanoic
acid (2-OH TCA) is characterized by the reversible appearance
and growth of large folds projecting
into the subphase\cite{ref9}. Folds can be observed by
light-scattering microscopy (LSM) -- which is sensitive to
height differences -- with the resulting images providing a
projection of the fold on the air-water interface. As shown in
Fig.~1b, the shape of folds near onset indeed can be
fitted to a semi-ellipse. The growth of a number of folds
upon area reduction was recorded on videotape,
with the simultaneous changes in the fold width $U(0)$ and the
fold length $L$, obtained from a fit of the fold to a
semi-ellipse, determined by a frame-by-frame analysis of the
images. Figure~2 shows $U(0)$ vs. $L$ for three representative
folds. As predicted, $U(0)$ depends linearly on $L$ for
each of the folds and the slopes $\Gamma$
are essentially identical, $0.25 \pm 0.5$.
The intercepts are zero within the (limited) precision of the data.

In order to compare the measured values of $\Gamma$ and that
predicted by Eq.~1,
we studied the rheological properties of 2-OH TC. From the
pressure-area isotherm, we obtained an area compression
modululus of $610 \pm 10\ {\rm mN/m}$ near
the collapse pressure. Using a Couette viscometer,
designed for rheology studies of LM's \cite{ref12},
we measured the frequency-dependent complex
shear modulus $G^{*}(\omega)$ near the collapse pressure over
a frequency range of 0.001
to 0.1 Hz. As shown in Fig.~3, the real (elastic) part $G'(\omega)$
of the shear modulus
varies between 0.4 and 7 mN/m over this frequency range,
while the imaginary (dissipative) part $G''(\omega)$ varies
between 1.6 to 16 mN/m. In another experiment
the relaxation of shear stress was monitored by rotating
the outer cylinder with a constant rate of strain of
$0.005\ {\rm s^{-1}}$ for 30 s while monitoring the stress on the
inner cylinder. The outer cylinder was suddenly stopped, and
the stress on the inner cylinder was measured for an additional
30 s. The resulting stress relaxation could be fitted by a
single relaxation time $\tau_s = (24 \pm 2)\ {\rm s}$. These
results indicate that 2-OH TCA has to be considered as a
2D viscoelastic material, probably due to a significant area
concentration of 2D structural defects such as
grain boundaries.

In order to describe fold formation in a self-adhering
viscoelastic sheet, we continued to use Eq.~1 but simply
replaced the shear modulus $G$ by
$|G^{*}(\omega_t)|$ with $t = 2\pi/\omega_t$ the age of the
fold. For a Maxwellian viscoelastic
material, with a single stress relaxation time $\tau_s$,
this procedure reproduces Eq.~1
in the short time limit while in the long time limit it leads to a
semi-elliptical {\it flow profile} $V(X) \approx U(X)/\tau_s$,
consistent with fold formation in a self-adhering
2D fluid sheet having a surface viscosity $\eta = G\tau_s$.
Using Fig.~3 to estimate
$|G^{*}(\omega_t)|$, with t of the order of $10^2\ {\rm s}$,
together with our earlier result
$W/2 - \gamma = 13 \pm 2\ {\rm mN/m}$, we obtain for $\Gamma$
a value of the order of 0.6. This is in rather
reasonable agreement with the observed range but a
precise comparison clearly requires a consistent description of
fold growth in self-adhering viscoelastic sheets.

If we accept the mapping between LM folds and GC's, then we can
use the extensive literature on GC's to describe the formation of
reversible folds. For a GC, the stress field diverges near
the crack endpoints as $1/r^{1/2}$, and
the divergence of the stress leads to a zone of plastic
deformation or fracture near the endpoints. For visco-elastic LM's,
we must expect similar plastic high stress zones at the endpoints,
except that the film may be buckled out of the
plane in order to reduce this stress. The stress divergence near
GC endpoints produces a ``crack-widening force''
$f(L) = \epsilon L$ proportional to the crack length
$L$ \cite{ref1}, which should be present as well for folds.
We indeed observed a steady lateral widening of the giant folds
when we reduced the LM area. This fold-widening force, is
counteracted by the fold {\it line tension} $\tau$, due the highly
curved edge that borders the fold (see Fig.~1A). Within the Helfrich
description of LM's \cite{ref14}, the fold line tension $\tau$ is
related to the LM bending energy $\kappa$ and the film
thickness $d$ by $\pi\kappa/2d$. Measurements of the
bending modulus $\kappa$ by diffuse x-ray
scattering \cite{ref15} on LM's with tail lengths similar to
2-OH TCA lead to values for $\kappa$ in the range of
200 - 500 $\rm{k_B T}$.

Because of the competition between the two forces, a fold with
$L > \tau/\epsilon$ will grow spontaneously while for
$L < \tau/\epsilon$ a fold will shrink. The critical fold
length $L^* = \tau/\epsilon$ and the nucleation energy
barrier $\Delta E = 1/2 \tau L^*$ can be obtained
from the corresponding expressions for GC's \cite{ref1}:
\begin{equation}
L^* = \frac{2}{\pi}\frac{\tau Y}{(W/2 - \gamma)^2}
\end{equation}
The parameters entering Eq.~2 are experimentally
accessible.
For a viscoelastic LM we replace $G$ by
$|G^{*}(\omega_f)|$ with $2\pi/\omega_f$ the
formation time of the fold (of the order of seconds or faster).

Because the nucleation barrier diverges at $W/2$ equal to $\gamma$,
Eq.~2 explains why the collapse of 2-OH TCA required significantly
higher pressures than the nominal onset value
$\gamma_w - \gamma_H = 53 \pm 4\ {\rm mN/m}$
as predicted by Eq.~1. Next, since
formation times are short compared to the stress relaxation time,
$|G^{*}(\omega_f)|$ is of
the order of 100 nN/m or higher. Equation 2 then predicts that
the critical fold length $L^*$ should be in the range of
100 {\AA} or longer and that the nucleation
energy barrier $\Delta E$  should be in the range of
$20\ {\rm k_B T}$ or higher. It follows that {\it fold
formation in solid or viscoelastic LM's cannot proceed
spontaneously but requires the presence of large,
pre-existing structural defects}. We were not able
to identify the nature of these defects with LSM.
Diamant et al. \cite{ref16} proposed that
for multi-component LM's, phase boundaries would be natural sites
for the nucleation of folds
(due to differences in spontaneous curvature between
phases).  This explanation cannot hold for monophasic materials,
like 2-OH TCA, but buckling along grain boundaries may play a
similar role.

We conclude by noting that, according to Eq.~2, out-of-plane
reversible folding should be sensitively dependent on the
in-plane rheological properties of the LM. If fold formation is
desirable, as in the case of lung surfactant, than this could
be achieved with additives such as cholesterol that increase the
fluidity of a surfactant layer, and hence reduce
$|G^{*}(\omega_f)|$ without affecting the
structural integrity of the film.

Acknowledgements: This work was supported in part by
National Science Foundation grants CH 0079311(C.M.K.)
and CTS 0085751 (M.D.). We would like to thank T. Witten
for helpful comments and Sam Shinder for assistance in
the analysis of the fold images.

\begin{figure}
\epsfxsize = 3.0in
\centerline{\epsffile{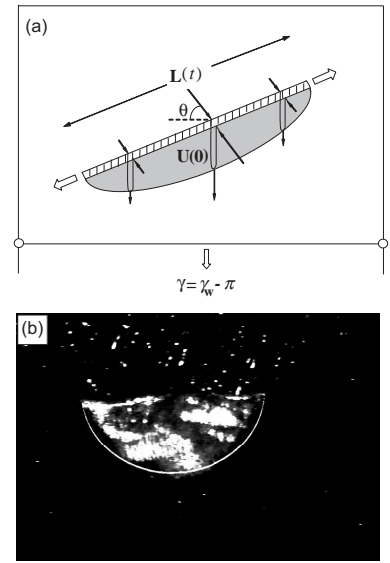}}
\caption{(a) Schematic diagram of a fold showing the
position-dependent fold width U(x) and the fold length L.
(b) Light scattering microscope image of an
early giant fold in 2-OH TCA. Its length is $65\ {\rm \mu m}$.
The fold is bent in a nearly
horizontal plane beneath the monolayer. The line is
a semi-ellipse fitted to the fold profile.}
\end{figure}

\begin{figure}
\epsfxsize = 3.0in
\centerline{\epsffile{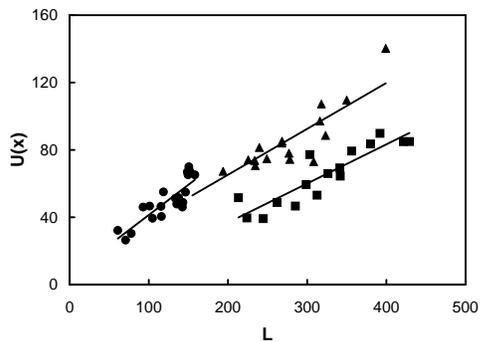}}
\caption{U(x=0) vs. L for three folds.  The data were
obtained from a frame-by-frame analysis of videotapes of
during fold widening under compression and are
given in numbers of pixels.}
\end{figure}

\begin{figure}
\epsfxsize = 3.0in
\centerline{\epsffile{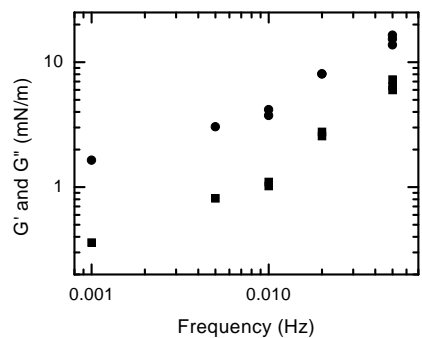}}
\caption{Values of $G'$ (lower line) and $G''$ (upper line)
in mN/m as a function of the oscillation frequency for
2-OH TCA at $20\ {\rm ^{\circ}C}$ at a pressure of 60 mN/m.}
\end{figure}

\end{document}